\documentclass[aps,pra,twocolumn,10pt]{revtex4-2}
\usepackage{graphicx}   % For including graphics
\usepackage{amsmath}    % For advanced math typesetting
\usepackage{amssymb}    % For additional symbols
\usepackage[colorlinks=true, linkcolor=blue, citecolor=blue, urlcolor=blue]{hyperref}  % Enable colored links
\usepackage{amsmath}
\usepackage[percent]{overpic}
\usepackage{subcaption}
\usepackage{natbib}
\usepackage{tikz}
\usepackage{xcolor}   % Enables color support
\usepackage{enumitem}
\newcommand{\highlight}[1]{\textcolor{black}{#1}}  % Define \highlight to apply red color
\begin{document}

\title{Ancilla measurement-based Quantum Otto engine using double-pair spin architecture}

\author{S. R. Rathnakaran}
\email{s.21phz0010@iitrpr.ac.in}
\author{Asoka Biswas}
\email{abiswas@iitrpr.ac.in}
\affiliation{Department of Physics, Indian Institute of Technology-Ropar, Rupnagar, Punjab, India}

\date{\today}  % Use \today to automatically insert the date or specify manually

\begin{abstract}

We present a quantum heat engine model utilizing a dual spin-pair architecture, wherein an Otto-like cycle is implemented using a single heat bath. The conventional cold bath is replaced by a measurement protocol, enabling engine operation without the need for a second thermal reservoir. Unlike standard quantum heat engines, our framework employs an ancillary spin pair in a two-dimensional configuration to regulate performance. Operating in finite time, the engine attains finite power, which is enhanced through quantum correlations, specifically correlation between spin pairs and projective measurements on the ancillary pair. The system consists of dual qubit pairs, where one pair serves as the working medium and the other as an ancillary system facilitating measurement-induced heat exchange. We demonstrate that the engine efficiency can exceed the standard quantum Otto limit through local control of the ancillary pair while maintaining nonzero power output. Moreover, correlation between spin pairs enables efficiency modulation via the measurement basis, underscoring the role of quantum resources in optimizing quantum thermal machines.

\end{abstract}

\maketitle

\section{Introduction}
Quantum thermodynamics is a rapidly developing field that extends classical thermodynamic concepts such as heat, work, and the laws of thermodynamics into the quantum regime \cite{doi:10.1080/00107514.2016.1201896,Binder:2018rix,10.1063/1.5096173}. Quantum thermal machines represent a growing frontier in the development of thermodynamic systems where quantum effects play a central role in energy transfer and information processing. Significant attention has been devoted to formulating quantum analogs of classical engines, leading to the development of various quantum heat engines (QHEs) such as Carnot \cite{Carl_M_Bender_2000,doi:10.1098/rspa.2001.0928,PhysRevE.73.036122}, Otto \cite{PhysRevE.85.041148,PhysRevE.79.041113,thomas2011coupled,PhysRevA.98.042102,PhysRevA.99.062103, PhysRevA.107.032220,PhysRevB.102.155423,PhysRevE.92.032124}, Brayton \cite{PhysRevE.87.012144,JMP}, Diesel \cite{JMP}, and Stirling\cite{PURKAIT2022128180,PhysRevE.103.062109}. These engines often operate on the basis of quantum versions of classical thermodynamic cycles, particularly the Otto and Carnot cycles. Numerous quantum systems have been proposed as working substances, including qubits \cite{e22070755}, qudits \cite{PhysRevE.92.032124,PhysRevE.92.042123}, photons \cite{science.1078955,Türkpençe_2017}, and harmonic oscillators \cite{PhysRevE.94.012137,PhysRevE.91.062134}. In addition, many-body quantum engines \cite{PhysRevB.99.024203,PhysRevLett.124.110606, PhysRevB.109.024310,PhysRevLett.124.170602,PhysRevA.108.042219} have also been explored. Experimental advances \cite{PhysRevLett.123.240601,10.1116/5.0083192,PRXQuantum.2.030353} now allow the construction of these QHEs, enabling the testing and verification of theoretical predictions. In particular, spin chains \cite{PhysRevB.101.180301,PhysRevB.107.085116,PhysRevB.107.235113,Piccitto_2022,Cavazzoni_2024,PhysRevA.110.012443,PhysRevB.107.075116,PhysRevE.92.032142,centamori2023spinchainbasedquantumthermal} have emerged as a promising platform for the implementation of such machines, due to their inherent ability to encode quantum information and facilitate controlled interactions between individual quantum systems. These machines operate under the principles of quantum thermodynamics, which extend classical thermodynamic concepts to systems where quantum coherence \cite{PhysRevA.99.062103,PhysRevE.99.042105,PhysRevE.102.062152,PhysRevResearch.2.033279}, entanglement \cite{PhysRevA.72.014301,PhysRevA.109.042422}, and non-equilibrium dynamics \cite{PhysRevA.102.050203,PhysRevE.109.034112} become significant. The use of spin chains in quantum thermal machines offers several advantages. Spin systems can be fine tuned through interactions with external fields, enabling precise control over energy levels and couplings. Furthermore, the interplay of spin-spin interactions, as well as coupling to thermal environments, leads to rich behavior that can be harnessed for quantum-enhanced tasks such as cooling, work extraction \cite{
PhysRevE.92.032142,PhysRevA.110.012443}, and heat rectification \cite{PhysRevE.102.062146,PhysRevLett.120.200603}. This makes spin chains ideal candidates for investigating fundamental questions in quantum thermodynamics as well as for designing practical quantum devices that perform specific thermodynamic tasks. 

Since Maxwell's time, it has been understood that work can be extracted from a single-temperature heat bath \cite{science.1078955,PhysRevLett.87.220601,PhysRevE.86.051105,PhysRevA.104.062210} using information obtained through measurements. This concept is embodied in Szilard’s engine, where feedback based on selective measurements \cite{PhysRevA.108.062214,PhysRevB.109.224203,PhysRevE.96.022108,PhysRevE.98.042122,PhysRevLett.122.070603} is employed to drive the engine’s operation. More recently, projective measurements of a system in its ground state have been shown to effectively simulate the release of heat to a cold bath \cite{PhysRevE.95.032111} during an isochoric process, highlighting the role of quantum measurement in thermodynamic processes. Finite-time analysis is crucial in the study of QHEs, as practical applications require the generation of finite power. Moreover, the performance of QHEs in finite-time operations may exhibit quantum characteristics that are not observable in quasistatic regimes \cite{PhysRevE.103.032144,PhysRevE.107.054110}. Although the effects of measurement processes and finite-time operation on QHE performance have been studied in different context and systems, there is a notable lack of research exploring the combined impact of these protocols.

In quantum phase transitions (QPTs), particle interactions exhibit long-range behavior governed by quantum fluctuations rather than thermal ones. A QPT represents a transformation in ground state properties at a critical point, the quantum critical point (QCP), where a non-thermal control parameter, such as field strength, reaches a critical value at absolute zero \cite{PhysRevE.98.052147}. At the QCP, quantum fluctuations significantly impact both finite-time and quasi-static thermodynamic cycles, distinguishing QPTs from thermal phase transitions, which occur at non-zero temperatures. In zero-temperature quantum systems at the thermodynamic limit, phase transitions emerge as nonanalytic behavior in ground-state energy derivatives when a Hamiltonian parameter is tuned. First-order QPTs show nonanalyticity via level crossings, while second-order QPTs reveal critical points at avoided crossings. %Such discontinuity is a generic feature of the  QPTs. 
Similar discontinuous behavior is also exhibited by the classical latent heat in classical phase transitions. This sudden change is not quantum reminiscent of classical latency, rather a novel form of non-equilibrium quantum latency \cite{PhysRevE.89.062103}. This universal behavior implies that a minimal model that features both types of crossings effectively captures QPT thermodynamics. 

In this study, we model a quantum Otto Engine (QOE) using a double-chain, or ladder, structure of qubits with XX (spin-flip-type) couplings between neighboring qubits along and between the chains (rungs). A pair of coupled qubits, defined as the `system', is driven by a time-dependent external magnetic field, while a second pair, coupled to this system, serves as an `ancilla'. This configuration undergoes a first-order QPT when the external magnetic field strength $B$ equals the spin-spin interaction strength $|J_1|$, causing a ground-state energy level crossing at low temperatures. In the range $-J_1 \leq B \leq J_1$, the ground state of the system is doubly degenerate, with critical points at $J_1 = \pm B$, designated as quantum critical points (QCP). To lift this degeneracy, we introduce small transverse fields $\delta_1$ and $\delta_2$ to the system and ancilla, respectively, emulating a Landau-Zener Hamiltonian \cite{PhysRevA.71.012307}. This perturbation results in non-commuting Hamiltonian at two different times, leading to transitions between instantaneous energy eigenstates and imbuing the QOE performance with quantum characteristics through finite-time unitary driving processes. Using the QPT, we demonstrate that the engine efficiency exceeds the Otto limit at the QCP. In addition, we emulated a cold bath using projective measurements on both the system and the ancilla. The standard approach applies measurements to the system connected to a heat bath at temperature $T$, whereas in the ancilla-assisted method, measurements are made on the ancilla without disturbing the system directly. Our results reveal that the ancilla-assisted measurement technique outperforms the standard measurement-based heat engine \cite{PhysRevE.107.054110}.

The structure of this paper is as follows. In Section II A, we introduce the Heisenberg 2-D XX spin chain (double chain) \cite{PhysRevA.94.032338,PhysRevB.65.104415,PhysRevE.92.022108,PhysRevE.109.014134} as the working model. Here, we describe the stages of the quantum Otto cycle, define the associated thermodynamic quantities, and analyze the QOE’s behavior near the critical point in finite-time regimes. In Section II B, we examine the influence of ancilla-based measurements on engine performance, focusing on the implementation of quantum Otto cycle operations. This framework also enables additional functionalities of the thermal machine, including refrigeration, heating, and acceleration. In Section III, we present a comparative analysis of both models, discuss the experimental feasibility of our approach, and conclude the findings in Section IV.

\section{2-D Heisenberg XX Spin-pair Models}
\subsection{System Measurement-based Engine}
\begin{figure}
    \centering
    \includegraphics[width=0.7\linewidth]{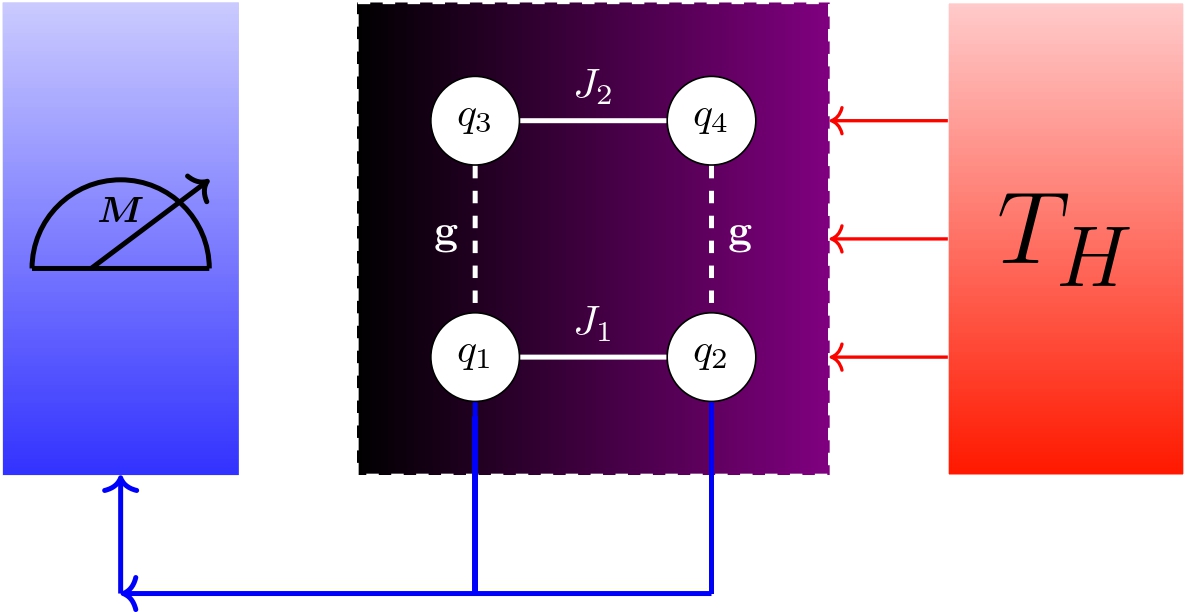}
    \caption{System based Measurement Engine}
    \label{fig:1}
\end{figure}

We consider a model (see Fig.~\ref{fig:1}), in which \{$q_{1}$, $q_{2}$\} represent the system qubits and \{$q_{3}$, $q_{4}$\} the ancillary qubits.                
The Hamiltonian of the entire system is given by 
\begin{equation}
\boldsymbol{H_{\text{tot}}} = \boldsymbol{H_{\text{sys}}} + \boldsymbol{H_{\text{anc}}} + \boldsymbol{H_{\text{int}}}\;,
\end{equation} 
where the system and ancilla Hamiltonians are given, respectively, as
\begin{equation*}
\begin{aligned}
    \boldsymbol{H_{\text{sys}}(t)} &= J_{1} \left(\sigma_{x}^{(1)} \sigma_{x}^{(2)} + \sigma_{y}^{(1)} \sigma_{y}^{(2)} \right) 
         + B(t) (\sigma_{z}^{(1)}+\sigma_{z}^{(2)}) \\
         &\quad + \delta_{1} (\sigma_{x}^{(1)}+\sigma_{x}^{(2)})\;, \\
    \boldsymbol{H_{\text{anc}}} &= J_{2} \left(\sigma_{x}^{(3)} \sigma_{x}^{(4)} + \sigma_{y}^{(3)} \sigma_{y}^{(4)} \right) 
         + \omega \left(\sigma_{z}^{(3)} + \sigma_{z}^{(4)}\right) \\
         &\quad + \delta_{2} \left(\sigma_{x}^{(3)} + \sigma_{x}^{(4)}\right)\;, \\
    \boldsymbol{H_{\text{int}}} &= g \left(\sigma_{x}^{(1)} \sigma_{x}^{(3)} + \sigma_{y}^{(1)} \sigma_{y}^{(3)} 
         + \sigma_{x}^{(2)} \sigma_{x}^{(4)} + \sigma_{y}^{(2)} \sigma_{y}^{(4)} \right)\;.
\end{aligned}
\end{equation*}
Here, the coupling strengths $J_{1}$ and $J_{2}$ denote the intrapair interactions of the system qubits and the ancillary qubits, respectively. The cases $J_{i} > 0$ ($J_{i} < 0$) correspond to antiferromagnetic (ferromagnetic) domain.  Furthermore, the system and the ancilla qubits are coupled with each other via an interpair coupling $g$. This configuration models a system-ancilla architecture, where intrapair and interpair couplings are crucial in the dynamics and information transfer between the qubits. Here $\hat{\sigma}_{i}^{x,y,z}$ denote the usual Pauli spin matrices for the $i^{th}$ spin ($i \in 1,2,3, 4$) and $\delta_{1}$ ($\delta_{2}$) is a time-independent transverse field applied along the transverse direction to the system (the ancilla). 
This transverse field makes the total Hamiltonian non-commuting at two different times:  $\bigl[H_{total}(t_{1}), H_{total}(t_{2})\bigr] \neq 0$, indicating possibility of quantum friction at finite times.  The external time-dependent magnetic field $B(t)$ is applied on the system qubits $q_{i}, (i \in 1, 2$) in a linear ramp: $B(t) = B_{L} + (B_{H} - B_{L})t/\tau$, where the magnetic field is changed from $B_L$ at $t=0$ to $B_H$ at $t=\tau$.

\subsubsection{\textbf{Quantum Otto Cycle}}
The combined system-ancillae, initially prepared in the state $\hat{\rho}_{A} = (|0 \rangle\langle 0|)^{\otimes 4}$, % which corresponds to the configuration 
%$|q_{1},q_{2},q_{3},q_{4}\rangle \langle q_{1},q_{2},q_{3},q_{4}|$.  
undergoes a unitary expansion as the magnetic field is swept from $B_{L}$ to $ B_{H}$  over a finite time interval, from $t=0 $ to $t=\tau$.  After this transformation, the evolved state of the system is given by $\hat{\rho}_{B} = \hat{U}(\tau) \rho_{A} \hat{U}(\tau)^{\dagger}$, where $\hat{U}(\tau) = \mathcal{T} \exp \left[ -i \int_0^{\tau} dt \, \hat{{H}}_{\text{tot}}(t) \right]$ represents the time evolution operator with explicit time ordering denoted by $\mathcal{T}$. %, and $\hat{{H}}_{\text{tot}}(t)$ is the total time-dependent Hamiltonian governing the dynamics. 
The work extracted during this process is given by
 \begin{equation}
 W_{1}(\tau)=\int_0^\tau \operatorname{Tr}\left[\hat{\rho}_{A\xrightarrow{}B}^{sys}\left(\tau^{\prime}\right) \dot{\hat{H}}_{sys}\left(\tau^{\prime}\right)\right] d \tau^{\prime}.
 \label{eq:work1}
\end{equation} 
After the unitary process, the system is coupled to a thermal bath at inverse temperature \( \beta_{H} = 1/T_{H} \), where the constants \( k_{B} \) and \( \hbar \) are set to unity. The bath drives the system to its thermal equilibrium state \( \hat{\rho}_{C}=\hat{\rho}_{th} \) during the isomagnetic stroke, with the magnetic field \( B_{H} \) held constant.
The heat absorbed by the system during this isomagnetic process is given by 
\begin{equation}
 Q_{H}=\int_0^{t_{h}} \operatorname{Tr}\left[\dot{\hat{\rho}}^{sys}\left(t\right) \hat{H}_{sys}\left(t\right)\right] d t.
 \label{eq:heat}
\end{equation} 
In the subsequent stage, the working system is decoupled from the thermal bath and undergoes a unitary compression as the external field is reduced from \( B_{H} \) to \( B_{L} \) over a finite time \( \tau \), governed by the protocol \( B(t - \tau) \). The state of the system at the end of this stroke is given by \( \hat{\rho}_{D} = \hat{U'}(\tau) \hat{\rho}_{C} \hat{U'}^{\dagger}(\tau) \), where \( \hat{U'}(\tau) = \mathcal{T} \exp \left[ -i \int_0^{\tau} dt \, \hat{H}_{\text{tot}}(t - \tau) \right] \) denotes the time-ordered unitary evolution operator.
The work extracted during this process is given by 
\begin{equation}
 W_{2}(\tau)=\int_0^\tau \operatorname{Tr}\left[\hat{\rho}_{C\xrightarrow{}D}^{sys}\left(\tau^{\prime}\right) \dot{\hat{H}}_{sys}\left(t-\tau^{\prime}\right)\right] d \tau^{\prime}.
 \label{eq:work2}
\end{equation} 

\highlight{
To complete the cycle, the working system undergoes cooling, which is emulated by selectively projecting the state $\hat{\rho}_D$ onto the state $|00 \rangle_{sys}\langle 00|$. %$(|00 \rangle_{anc}\langle 00|)$ onto the qubit pair \(\{q_1, q_2\}\) \((\{q_3, q_4\})\).
via  a measurement operator given by $\hat{M} = |00\rangle_{sys}\langle 00| \otimes I_{anc}$, where $\hat{\rho}_{D}$ is the density matrix of the system + the ancilla prior to measurement and $I_{anc}$ is the identity operator in the ancilla basis. 
The post-measurement state of the system can be written as $\hat{\rho}_{\text{PM}} = \hat{M} \hat{\rho}_{D} \hat{M}^{\dagger}/p_m$,
where $p_m = \operatorname{Tr}[\hat{M} \hat{\rho}_{D} \hat{M}^{\dagger}]$ is the probability of measuring the system into $|00\rangle$ state. For our protocol, we assume $p_m=1$, i.e., we consider only those measurement events where the outcome is certain.}

%\begin{equation}\highlight{
  %   = |00\rangle_{\text{sys}}\langle 00| \otimes \hat{\rho}_{\text{anc}|00} } ,
%\end{equation}

%\highlight{Case II:  Measurement done on ancilla qubits.}
%\begin{equation} \highlight{
 %   \hat{\rho}^{\text{PM}}_{sys|00} = \frac{\hat{M} \hat{\rho}_{D} \hat{M}^{\dagger}}{p_n} = \hat{\rho}_{\text{sys}|00} \otimes |00\rangle_{\text{anc}}\langle 00|}    ,
%\end{equation}

%\highlight{, ($I_{sys} \otimes |00\rangle\langle00|$) is the measurement operator corresponding to the system (ancilla) and  The operator $\hat{M}$ is Hermitian, satisfying $\hat{M} = \hat{M}^{\dagger}$, and }

 The amount of heat lost during this process is given by 
\begin{equation}
    Q_{C} = \operatorname{Tr}[\hat{\rho}_{PM} \hat{H}_{sys} - \hat{\rho}_{D}^{sys} \hat{H}_{sys}]\;.
\end{equation}

\subsubsection{\textbf{QOE's quasi-static operation}}
The work strokes $W_{1}(\tau)$  and  $W_{2}(\tau)$ are performed over an extended cycle time $ \tau $, such that transitions between energy levels are effectively suppressed, rendering the driving process adiabatic. In this quasi-static regime, the efficiency of the cycle is expressed as 
\begin{equation}
    \eta = 1 - \frac{B_{L}}{B_{H}}\;.
    \label{Eq:7}
\end{equation}

\begin{figure}[h]
    \centering
    \includegraphics[width=1.1\columnwidth]{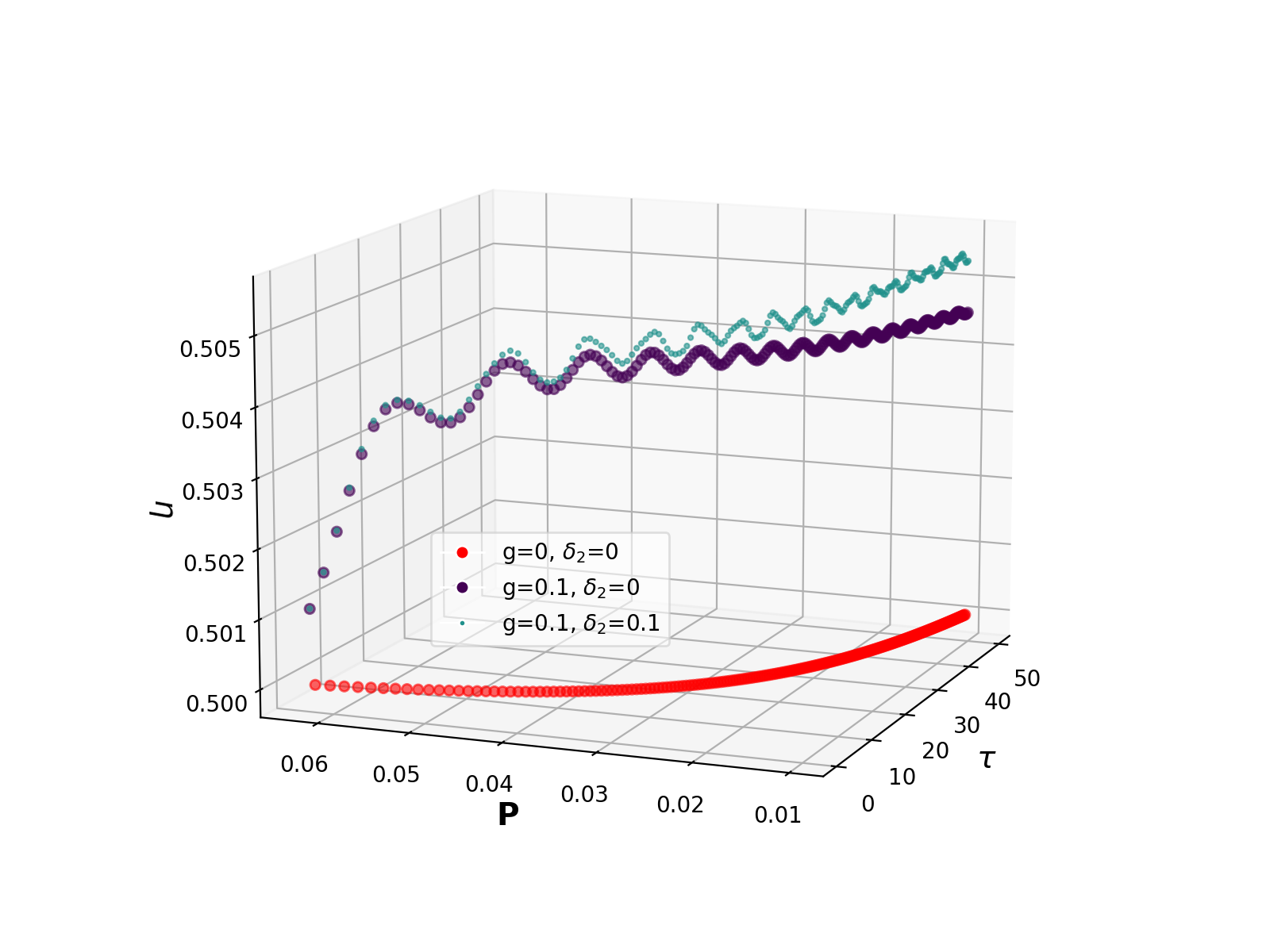}
    \caption{Comparison of efficiency vs power dynamics under different conditions. Given the system parameters: $T = 10$, $J_{1} = J_{2} = 1$, $\delta_{1} = 0$, $B_{L} = 1$, $B_{H} = 2$, $\omega = 0.5$. All parameters are normalised with $J_{2}$.}
    \label{fig:2}
\end{figure}

The Fig.\ref{fig:2} illustrates the variation of efficiency $\eta$ and power (\textbf{P}) as a function of the unitary process time $\tau$. It is observed that the variation of $\eta$ exactly follows the Eq.\ref{Eq:7},  when the ancilla are decoupled from the system. i.e., when $g = 0$. But the power \textbf{P}  deteriorates in the quasi-static limit ($\tau \xrightarrow{} \infty $)  as expected.

\highlight{
In contrast, when the ancillary system is coupled with the system (i.e., when $g\ne 0$) as shown in Fig. \ref{fig:1}, there is a noticeable improvement in efficiency. This enhancement becomes more pronounced for larger $\delta_{2}$ in the quasi-static limit $\tau \xrightarrow{} \infty $, and it becomes clear that a local control field in the ancilla can lead to greater efficiency than the Otto efficiency of a single qubit given in \eqref{Eq:7}, which serves as a robust benchmark and is restored when the interaction between the system and the ancilla is turned off ($g = 0$) as evidenced by the constant efficiency (flat line) output in Fig.~\ref{fig:2} which is consistent with the XX model as discussed \cite{PhysRevE.107.054110}.
}

\subsubsection{\textbf{Critical point behaviour}}

\begin{figure*}[tb]
    \centering
    
    \begin{subfigure}[b]{0.48\textwidth}
        \centering
        \begin{tikzpicture}
            \node (img) {\includegraphics[width=0.8\textwidth]{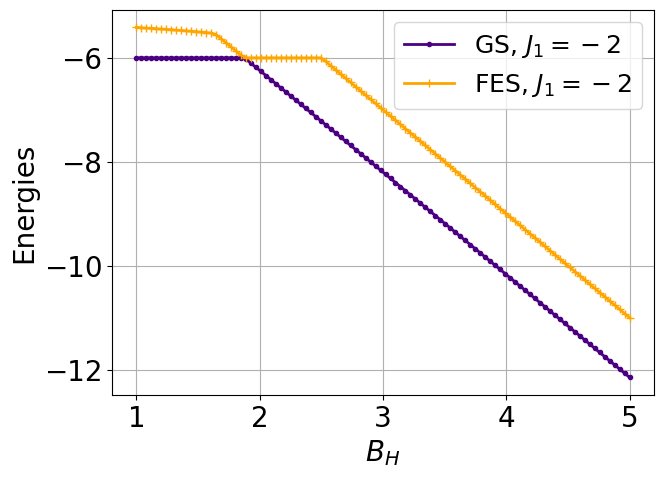}};
            \node at (img.south east) [anchor=south east, xshift=-2cm, yshift=1.0cm] 
            {\includegraphics[width=0.4\textwidth]{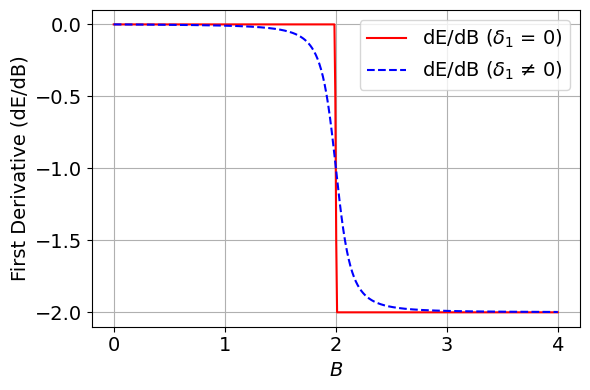}}; % Inset image
        \end{tikzpicture}
        \caption{First order QPT.}
        \label{fig:1qpt}
    \end{subfigure}
    \hfill
    \begin{subfigure}[b]{0.48\textwidth}
        \centering
        \begin{tikzpicture}
            \node (img) {\includegraphics[width=0.8\textwidth]{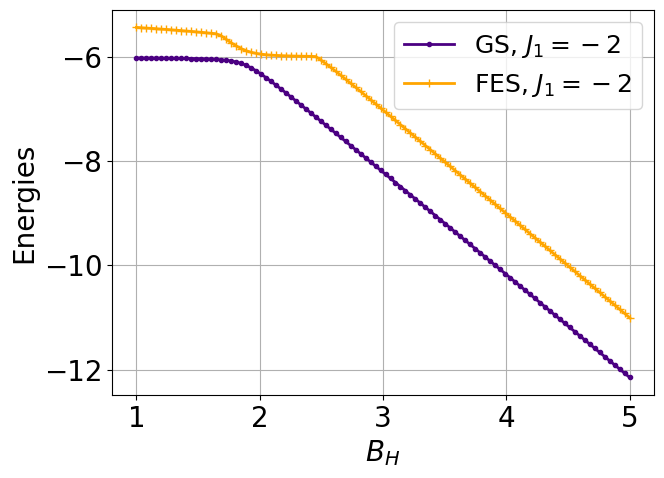}};
            \node at (img.south east) [anchor=south east, xshift=-2cm, yshift=1.0cm] 
            {\includegraphics[width=0.4\textwidth]{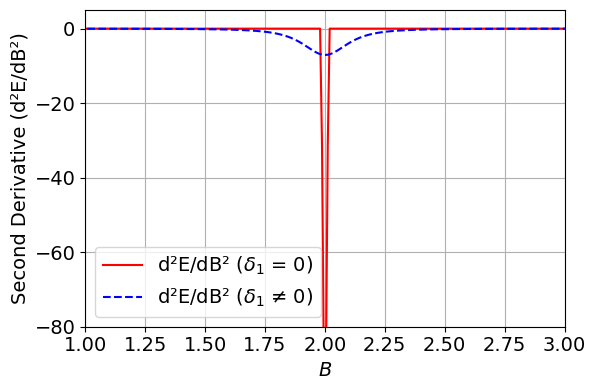}}; % Inset image
        \end{tikzpicture}
        \caption{Second order QPT.}
        \label{fig:2qpt}
    \end{subfigure}
    
    \caption{Energy level analysis of the ground and first excited states in the context of Landau-Zener Hamiltonians,
plotted as a function of the system’s control parameter $B_H$. (a) When $\delta_1$= 0 , a first-order quantum phase transition (QPT) occurs precisely at the level crossing, where the ground state (gs) and first excited state energies (fes) intersect discontinuously. 
 (b) For $\delta_1 \neq 0$ , an avoided level crossing is observed,
indicating a second-order QPT, where the transition is continuous, and the energy gap remains finite. Inset: plots for the first and second order derivatives of energy eigenvalues as a function of control parameter $B_H$. System parameters are same as that of in Fig. \ref{fig:2}}
    \label{fig:qpt}
\end{figure*}

\highlight{Quantum phase transitions are classified based on the non-analytic behavior of the ground-state energy:	A first-order QPT occurs when the first derivative of the ground-state energy is discontinuous.	A second-order QPT occurs when the second derivative is discontinuous while the first derivative remains continuous. For $\delta_1 = 0$, the system undergoes a first-order transition at $B \approx J_1$ due to level crossing, causing a discontinuity in $dE/dB$ (see the inset in Fig. \ref{fig:1qpt}). For $\delta_1 \neq 0$, the level crossing is replaced by an avoided crossing, smoothing out the first derivative. However, the second derivative remains finite and well-behaved for finite $\delta_1$, but it becomes singular in the limit $\delta_1 \to 0$ (see the inset in Fig. \ref{fig:2qpt}). This is the hallmark of a second-order QPT. Thus, the system undergoes a second-order QPT in the limit $\delta_1 \to 0$, as the avoided crossing turns into a sharp kink in the ground-state energy.}
We investigate the critical point behavior of the QHE in both the  $J_1 < 0 $ (ferromagnetic) and  $J_1 > 0$  (antiferromagnetic) regimes, where critical behavior arises due to QPTs.\cite{PhysRevE.89.062103}.

\begin{figure*}[tb]
    \centering
    
    % First subfigure
    \begin{subfigure}[b]{1\columnwidth} % Adjust width to fit two subfigures side by side
        \centering
        \includegraphics[width=0.8\textwidth]{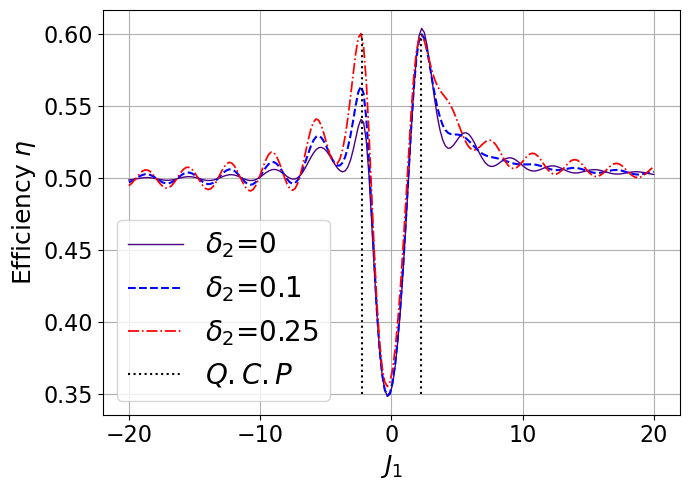} % Main figure
        \caption{Efficiency vs $J_{1}$}
        \label{fig:3}
    \end{subfigure}%
    %\hfill
    % Second subfigure
    \begin{subfigure}[b]{1\columnwidth} % Adjust width for second figure to fit beside the first
        \centering
        \includegraphics[width=0.8\textwidth]{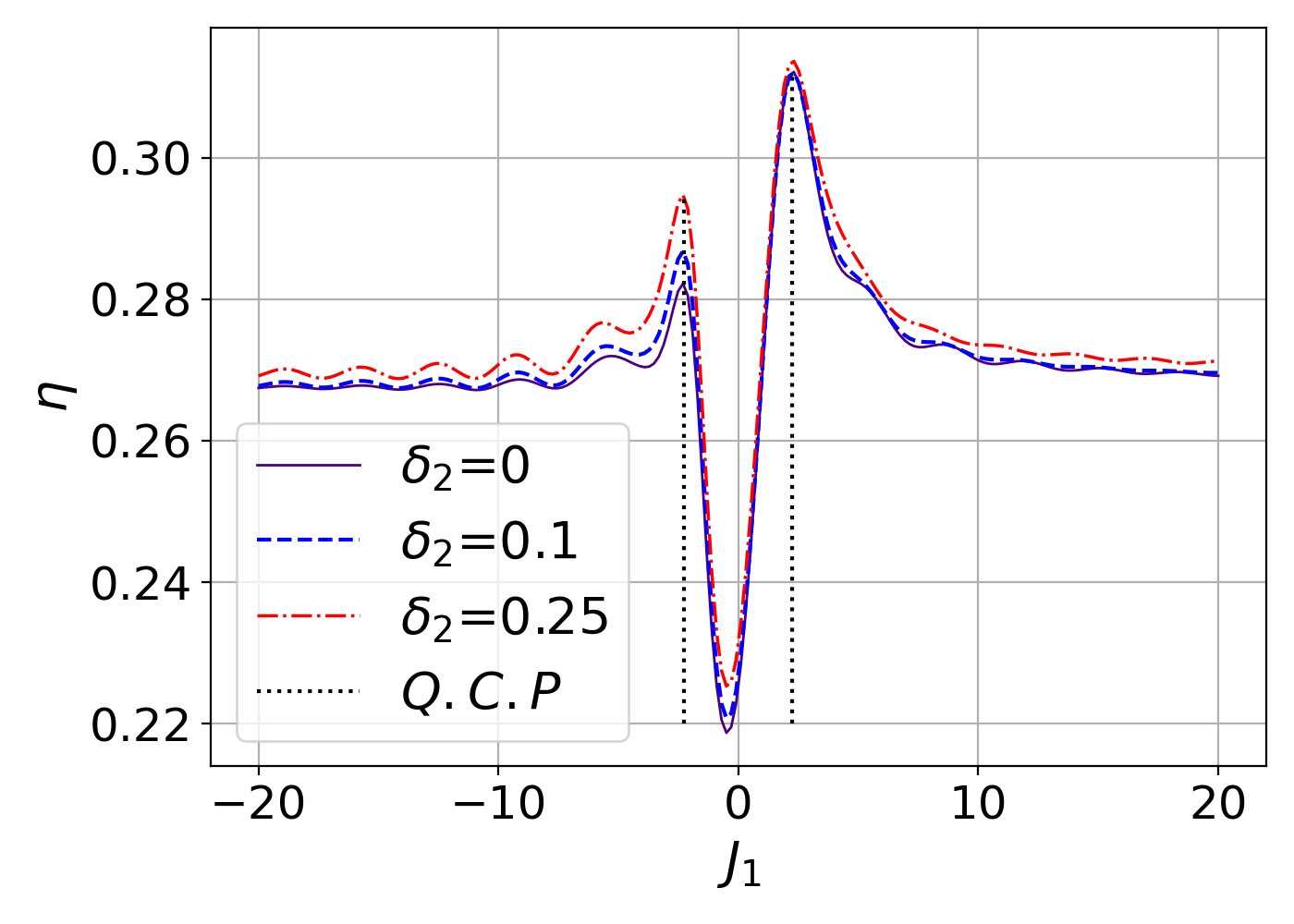} % Replace with your image file
        \caption{Efficiency vs $J_{1}$ for \textit{'always on interaction'}}
        \label{fig:3b}
    \end{subfigure}
    
    \caption{Efficiency vs $J_{1}$ for different values of $\delta_{2}$, $\delta_{1} = 1$, $\tau = 1$ and other system parameters are same as that of in Fig. \ref{fig:2}. The peak enhancement in the efficiency is due to the second-order QPT as illustrated in fig. \ref{fig:2qpt} }
    \label{fig:Fig.3 inset}
\end{figure*}

Fig. \ref{fig:3} illustrates the behavior of efficiency at QCPs in both antiferromagnetic (AFM) and ferromagnetic (FM) regimes. The efficiency peaks observed at the QCPs correspond to QPTs. At these critical points, level crossing occurs when the energy levels of the system Hamiltonian,  $\hat{H}_{s}$ , coincide, specifically for $  \delta_{1} = 0$, which refers to a first-order QPT. characterized by a discontinuity in the first-order derivative of energy eigenvalues  with respect to the magnetic field. However, for $\delta_{1} \neq 0$, an avoided crossing appears, which corresponds to a second-order phase transition, characterized by a continuous first-order derivative. However, the second-order derivative also remains finite and well-behaved for finite $\delta_1$, which would otherwise become singular in the limit $\delta_1 \to 0$.

\subsubsection{\textbf{Always on Interaction}}

The process described above represents the ideal scenario for the QOE, where perfect isolation from the bath is assumed during the work strokes.  However, in practical implementations, there is an inherent cost associated with coupling and decoupling the system from the bath. To address this issue and enhance the experimental feasibility of the system architecture, the bath remains coupled during the work strokes, resulting in a dissipative system. This dissipation must be taken into account when performing the work strokes. Consequently, solving the master equation with a time-dependent Hamiltonian in the presence of a dissipative bath becomes essential. Importantly, the measurement process is assumed to occur instantaneously, such that the bath does not influence the system during measurement. The Gorini–Kossakowski–Sudarshan–Lindblad (GKLS) master equation used here to simulate this dissipative process during the work strokes and also during the isomagnetic strokes to heat up the system where there are individual baths at temperature $T$ associated to each qubit is given by
\begin{widetext}
\begin{equation}
  \frac{\partial \rho}{\partial t} = \dot{\iota} [\rho, H] + \sum_{i} \left( \Gamma_{i}(t) \left( n_i(\omega_{t}) + 1 \right) \left( X_i \rho X_i^{\dagger}-\frac{1}{2} \left\{ X_i^{\dagger} X_i, \rho \right\}\right)\right)  +\Gamma_{i}(t) n_i(\omega_{t}) \left( X_i^{\dagger} \rho X_i - \frac{1}{2} \left\{ X_i X_i^{\dagger}, \rho \right\} \right).  
  \end{equation}
\end{widetext}

where $n\left(\omega_i\right)=\left[\exp \left(\frac{\hbar \omega_i}{k T}\right)-1\right]^{-1}$ is the average number of photons in the bath at the transition frequencies $\omega_i$. The $\Gamma_{i}(t) \left(n_i(\omega_{t})+1\right) $  and $\Gamma_{i}(t) n_i(\omega_{t})$  are the time dependent dissipation rates. We have chosen an Ohmic type spectrum of the bath coupling, $\Gamma_{i}(t) = 0.1\omega_{i}(t) e^{-\omega_{i}(t)/\omega_{c}}$, where $\omega_{c}$ is the cut-off frequency of the bath spectral density.

Note that the jump operators associated with a system decay are given by  
\begin{equation}
X(\omega)=\sum_{\epsilon^{\prime}-\epsilon=\omega}|\epsilon\rangle\left\langle\epsilon\left|X\right| \epsilon^{\prime}\right\rangle\left\langle\epsilon^{\prime}\right|,
\end{equation}
where $\{|\epsilon\rangle\}$ is the basis of the eigenvectors of the system Hamiltonian $H_s$.

\begin{figure}
    \centering
    \includegraphics[width=0.9\columnwidth]{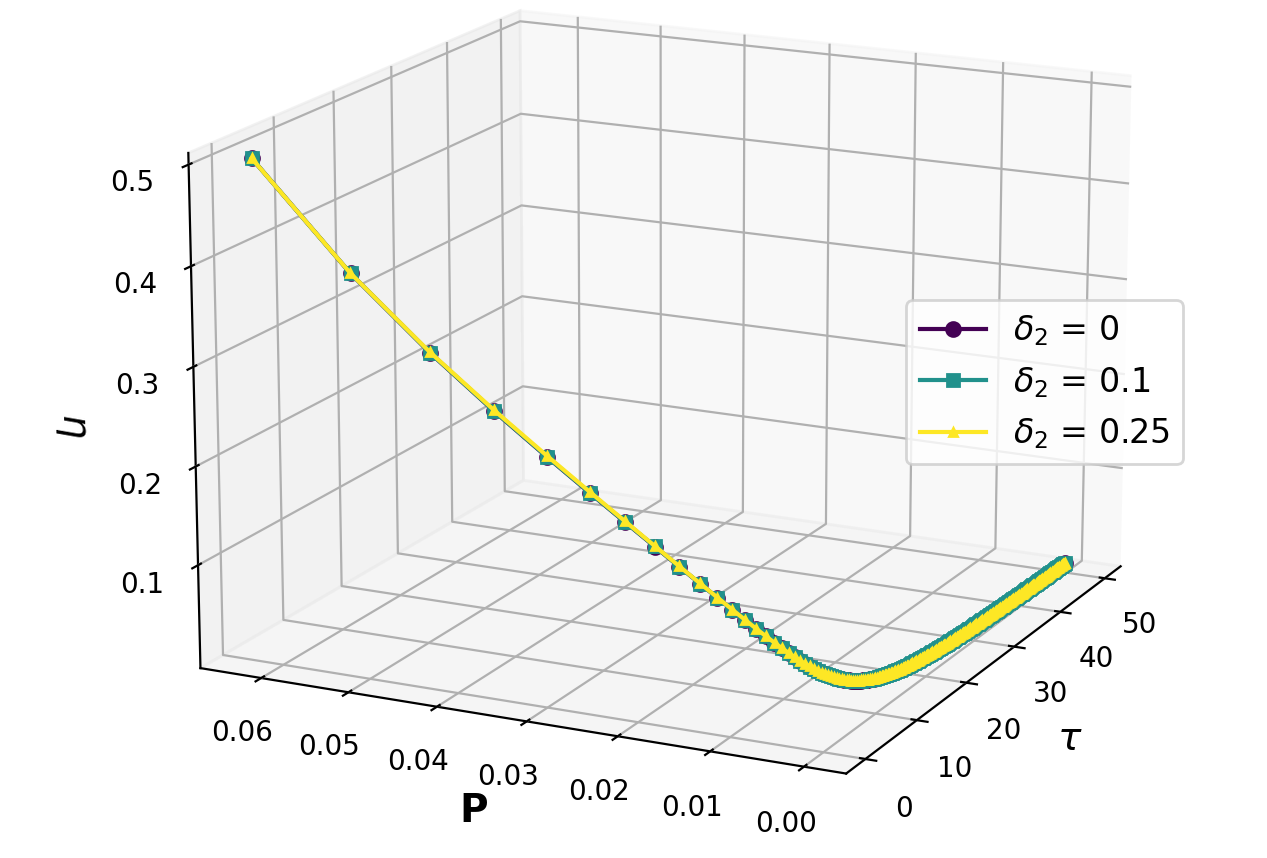}
    \caption{Efficiency vs Power dynamics in '\textit{always on interaction}'. The parametric values are same as  in Fig.\ref{fig:2}.}
    \label{fig:4}
\end{figure}

The damping of efficiency and power in QOEs, independent of the engine tuning parameter $\delta_2$, is observed when the bath effects are not deactivated during work strokes. This phenomenon is illustrated in Fig. \ref{fig:4}.
During finite-time operations, the engine efficiency corresponds to the standard Otto cycle efficiency limit as described by (\ref{Eq:7}). However, over extended periods, the system transitions to a steady state during the initial work stroke. This transition renders the isomagnetic heating stage negligible in terms of its impact on the overall system evolution. Consequently, persistent interaction with the bath (\textit{always on} interaction) consistently degrades the performance of the QOE over long durations.

\subsection{Ancilla Measurement-based Engine}

 \begin{figure}
    \centering
    \includegraphics[width=0.7\linewidth]{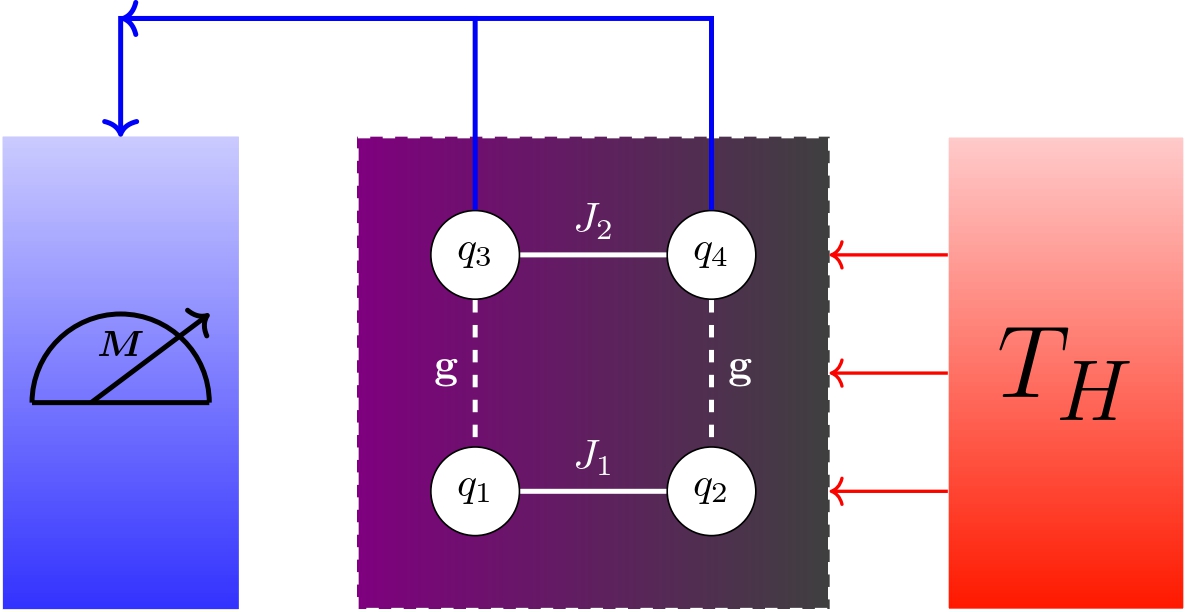}
    \caption{Ancilla based Measurement engine}
    \label{fig:5}
\end{figure}

In contrast to the conventional measurement on the system’s density matrix, $\hat{\rho}_{sys}$, typically employed to simulate the behavior of a cold bath (as shown in Fig.~\ref{fig:1}), we measure the ancillary qubits, $\hat{\rho}_{anc}$, as illustrated in Fig.\ref{fig:5}. This approach is advantageous because it avoids perturbation of the system qubits during the evolution, with measurements performed exclusively on the ancillae, $\hat{\rho}_{anc}$, which is coupled to the system by the intracoupling strength $g$. This nonzero coupling between system and ancillae creates a mutual exchange of energy. \\

\begin{figure*}[tb] % Places at top/bottom while keeping flow
    \centering
    \begin{subfigure}[b]{0.4\textwidth}
        \centering
        \includegraphics[width=\linewidth]{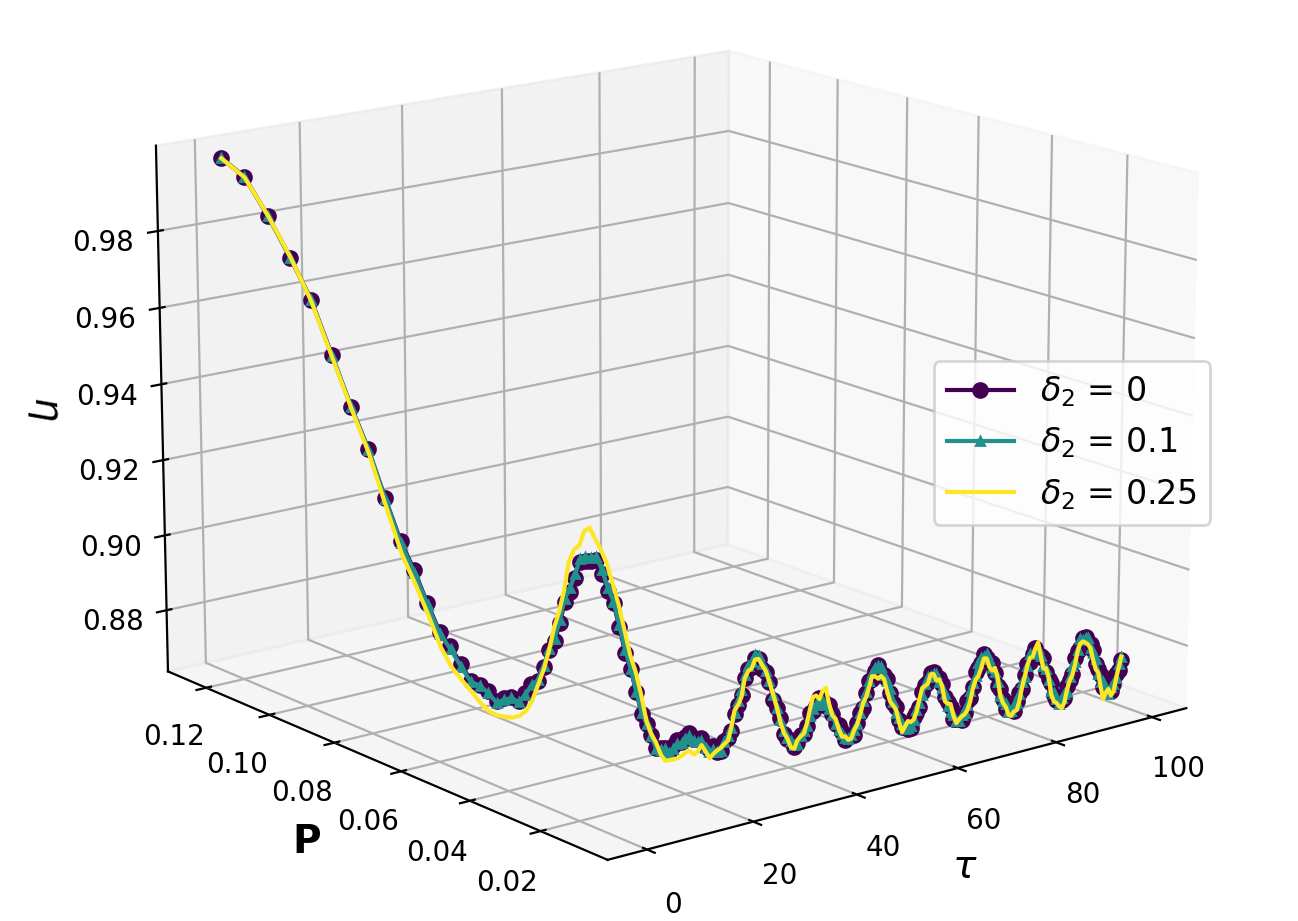}
        \caption{}
        \label{fig:6a}
    \end{subfigure}
    \hspace{0.12\textwidth}
    \begin{subfigure}[b]{0.37\textwidth}
        \centering
        \includegraphics[width=\linewidth]{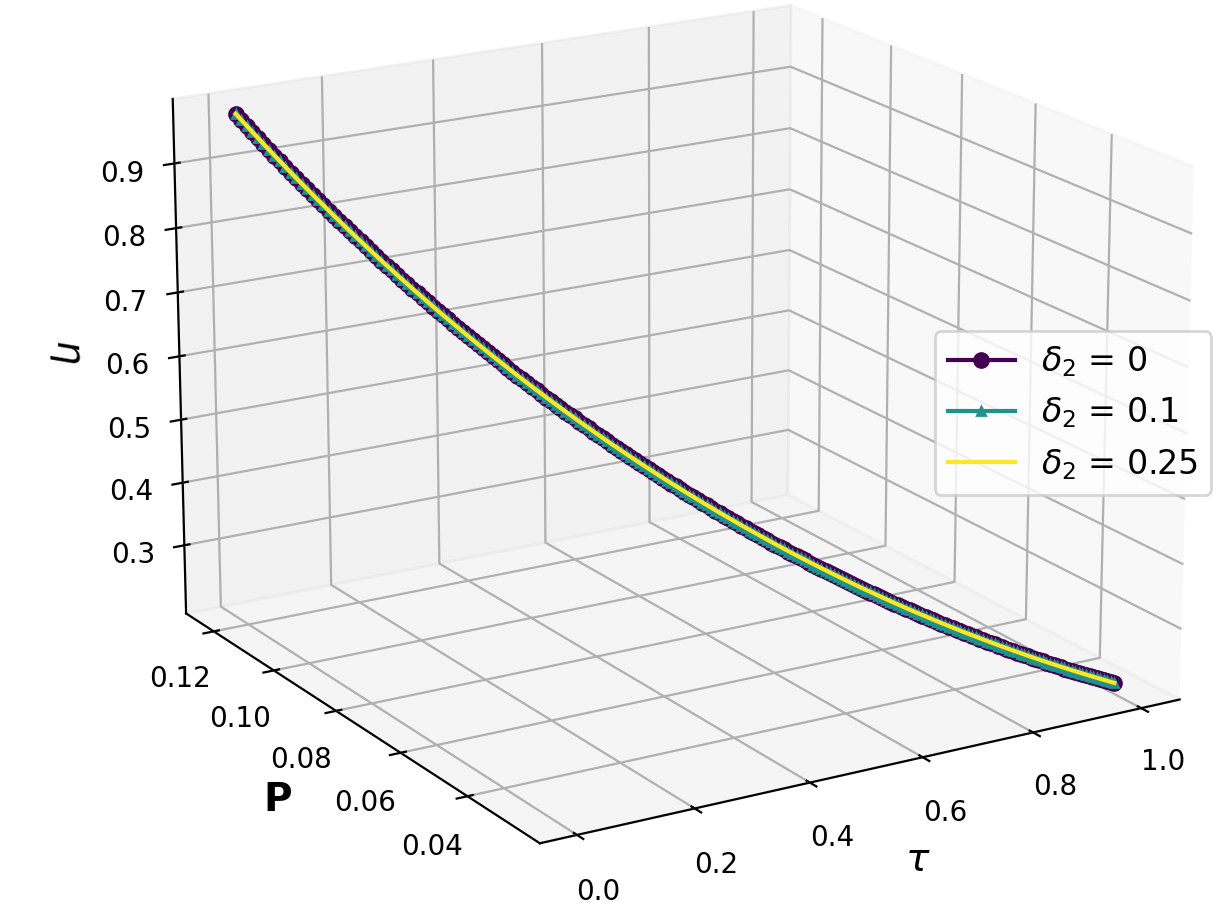}
        \caption{}
        \label{fig:6b}
    \end{subfigure}
    
    \caption{Comparison of efficiency vs power dynamics for standard ancilla-based QOE. (a) Efficiency vs Power dynamics in ancilla-based measurement. $J_{1}=2$, all other parametric values are the same as in Fig.\ref{fig:2}. (b) Efficiency vs Power dynamics with always-on interaction.}
    \label{fig:6}
\end{figure*}

Fig.\ref{fig:6a} depicts the variation of efficiency and power as a function of  $\tau$  for ancilla-based measurements. The plot reveals that both efficiency and power are enhanced in the finite-time regime compared to direct system measurements. \highlight{Such enhancement can be attributed to the non-zero transition probabilities among various energy eigenstates in the non-adiabatic timescale (see Appendix \ref{app:transitions})}.
The oscillatory behavior persists over extended timescales for different values of  $\delta_2$. However, in contrast to standard measurement results, an increase in  $\delta_2$  leads to a decrease in the QOE efficiency in this case.

\begin{figure}[ht]
    \centering
    \includegraphics[width=0.48\textwidth]{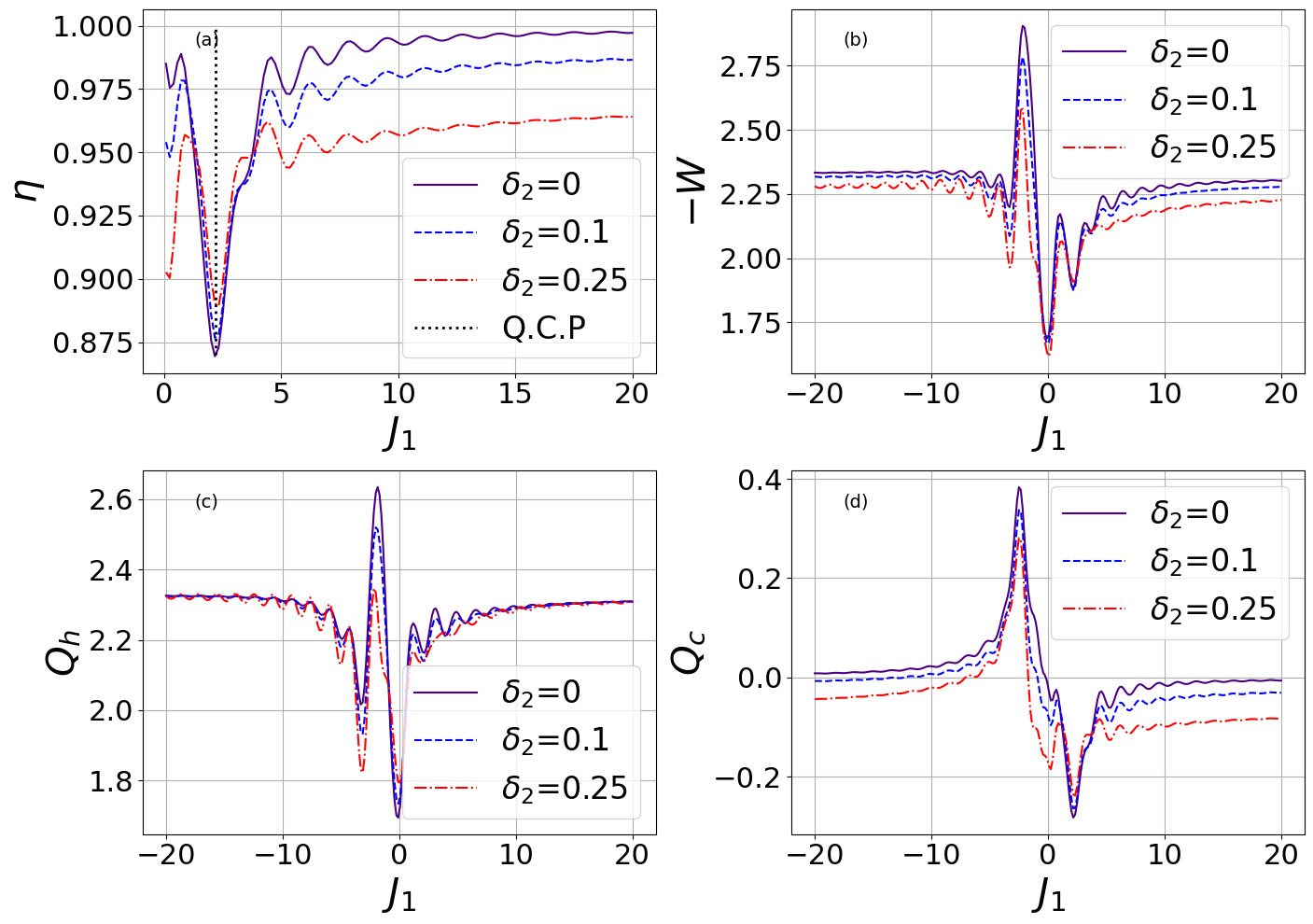}
    \caption{(a) Efficiency vs coupling strength $J_{1}$ for AFM interaction. (b) The total work done during the complete cycle $W$. (c) Heat exchange with the hot thermal bath at temperature $T$ ($Q_{H}$). (d) The energy exchange during the measurement stroke as the cold bath ($Q_{c}$). The parameters chosen are $g=0.75$, $\delta_{1} = 1$ and all other values are same as that of in Fig.\ref{fig:2}.}
    \label{fig:7a}
\end{figure}

\begin{figure}[ht]
    \centering
    \includegraphics[width=0.4\textwidth]{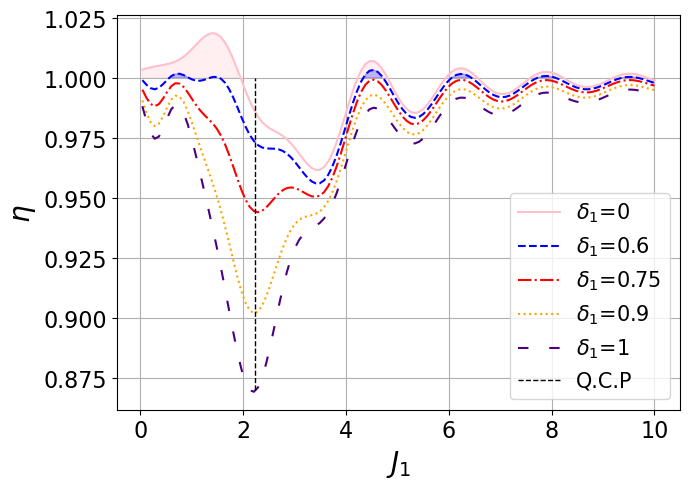}
    \caption{Efficiency vs $J_{1}$ for different values of $\delta_{1}$ where the shaded region shows the unphysical regime for AFM interaction. The parameters chosen are $\delta_{2} = 0$, all other parameters are same as that of in Fig.\ref{fig:2}.}
    \label{fig:7b}
\end{figure}

The thermodynamic quantities, including the work done (\(W\)), heat absorbed from the hot thermal bath at temperature \(T\) (\(Q_H\)), and heat rejected during the projective measurement on the ancillary part of the system (\(Q_C\)), are plotted as a function of the system coupling parameter \(J_1\) in Fig.\ref{fig:7a}. The work done is calculated as \(W_{\text{tot}} = Q_h - |Q_c|\), rather than the conventional \(W_{\text{tot}} = W_1 + W_2\), to account for the impact of the projective measurement on the ancillary subsystem. This alternative formulation accommodates the role of the measurement in influencing the overall energy exchange within the system. It is observed that the thermodynamic quantities adhere to the appropriate sign conventions within the regime of the QHE when the system coupling parameter satisfies \(J_1 > 0\) (AFM coupling). \highlight{This is especially evident during the measurement stroke, where \(Q_c < 0\), ensuring a physically consistent heat rejection process. For \(J_1 < 0\), the sign conventions deviate, leading to nonphysical results, indicating that the QHE operates meaningfully only in the AFM regime. A plot of the heat absorbed from the cold reservoir $Q_c$  vs $J_1$ reveals that in the regime where $ J_1 < 0$, the system absorbs heat ($Q_c > 0$). However, in this same regime, the other thermodynamic quantities—specifically, the total work done by the system ($W < 0$) and the heat input ($Q_h > 0$)—do not correspond to any of the four well-defined thermodynamic operation modes, namely \textbf{Engine, Refrigerator, Heater, or Accelerator}. This inconsistency suggests that the observed thermodynamic behavior in the $Q_c > 0$ regime is un-physical or does not correspond to a realizable thermodynamic cycle.}

The efficiency $\eta$ is plotted exclusively for the AFM regime, where it demonstrates enhanced performance compared to the standard measurement-based QHE. However, unlike the behavior observed at the QCP, there is a noticeable dip in efficiency. Additionally as the parameter \(\delta_2\) increases, the efficiency decreases, exhibiting behavior that is in stark contrast to the trend shown in Fig.\ref{fig:3}. This highlights the nuanced dependence of efficiency on system parameters in the AFM regime.

\subsubsection{\textbf{Many-cycle operation}}
\begin{figure*}
     
     \includegraphics[width=0.9\linewidth]{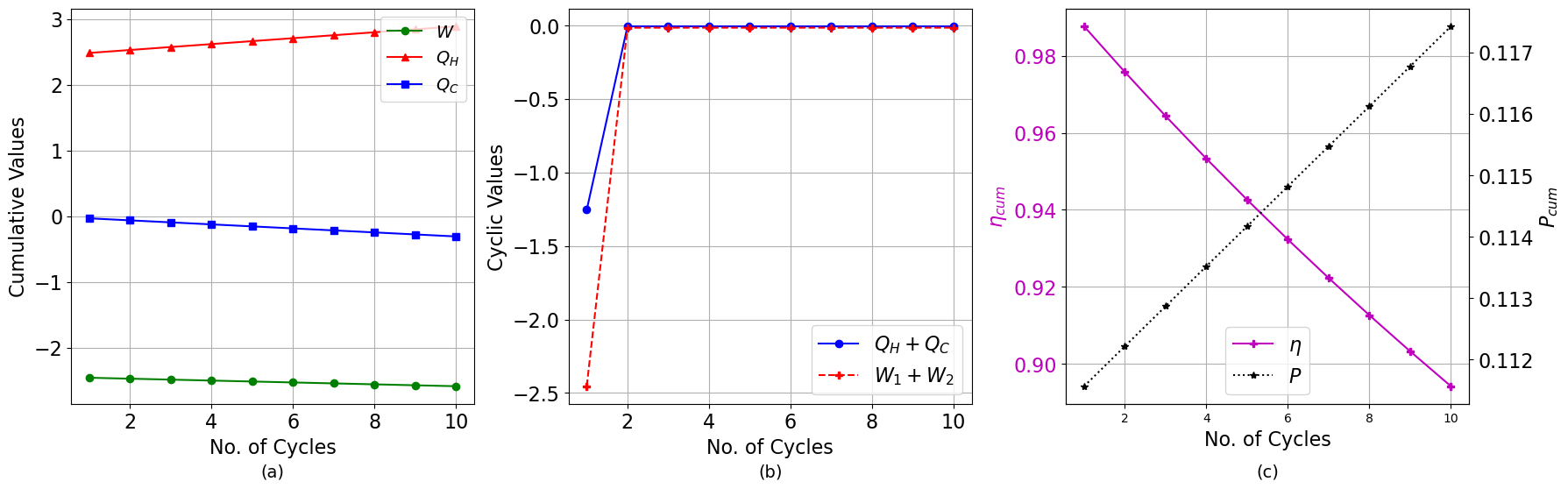}
     \caption{(a) The panel displays the cumulative values of work \(W\) (green circles), heat absorbed from the hot bath \(Q_H\) (red triangles), and heat rejected \(Q_C\) (blue squares) over multiple cycles. (b) This graph demonstrates the first law of thermodynamics, with \(Q_C + Q_H\) (blue circles) and \(W\) (red dotted crosses) converging to the same value at the QHE's limit cycle, confirming energy conservation. (c) The left axis shows the cumulative efficiency (magenta crosses) as a function of the number of cycles, while the right axis indicates the power output of the engine (black dotted stars). The parameters used are \(J_1 = 2\), \(\delta_2 = 0\), and \(\tau = 1\), with all other parameters as specified in Fig. \ref{fig:2}.}
     \label{fig:8}
\end{figure*}

Thus far, our analysis has focused on a single cycle of the engine. However, in practical applications, the engine operates over multiple cycles, which holds significant importance for understanding its long-term performance. Moreover, the engine's behavior in finite-time operations is crucial for assessing its real-world applicability, as the dynamics across multiple cycles and finite-time effects can considerably impact overall efficiency and power output. In the ancilla-based measurement protocol, the final state of the system ($\hat{\rho}_{\text{sys}|00}$) after the projective measurement, \( I_{sys}\otimes|00\rangle_{anc}\langle00| \), does not coincide with the system's initial state, \( |00\rangle_{sys}\langle00| \), preventing the cycle from closing. This discrepancy disrupts the consistency with the first law of thermodynamics, as illustrated in Fig. \ref{fig:8}b. 

However,  once we consider the engine operation for many cycles,  this discrepancy is resolved (\highlight{described in the next sub-section; [see Fig. \ref{Tracedist}]}), and the work calculated by both formulations aligns, ensuring that the first law is satisfied. The energies such as \(W_{\text{tot}} ,Q_h ,  Q_c \) are calculated cumulatively (Fig.  \ref{fig:8}a), The system operates as a heat engine, absorbing heat from the hot bath (\(Q_H > 0\)), expelling a portion of it to the cold bath (\(Q_C < 0\)), and generating useful work (\(W < 0\)). This thermodynamic process reflects the standard operation of a heat engine, where heat is converted into work while some energy is released to the cold reservoir. The cumulative value of efficiency over all cycles is given by  
\begin{equation}
    \eta_{cum} = \frac{\sum_{i} W_{tot}^{(i)}}{\sum_{i} Q_{in}^{(i)}}
\end{equation}
 as indicated in Fig. \ref{fig:8}c. The average efficiency over many cycles decreases, this is due to the minimal work done compared to the heat taken in by the system. The power developed over many cycles are important in the context of finite time dynamics. The power of the QHE can be defined as
 \begin{equation}
     P_{cum} =  \frac{\sum_{i} W_{tot}^{(i)}}{2\tau + t_{h}}
 \end{equation}
where $t_{h}$ is the isochoric heating time during the system contact with the bath. As the measurement is instantaneous there is no time involved during that stroke. In most heat engine models, there is typically a trade-off between efficiency and power. To achieve maximum power output, engines often operate well below the quasi-static efficiency limit. In short-time operations, which are necessary for producing high power, various types of irreversibilities lead to a reduction in both work output and efficiency. Conversely, in long-duration operations, while work and efficiency tend to improve, the extended timescale results in a significant reduction in power. This trade-off highlights the inherent challenges in optimizing both efficiency and power simultaneously. Although the efficiency is pronounced in the single cycle case, we can further optimize the efficiency-power trade-off without significantly compromising the performance of the QHE. This optimization allows for enhanced efficiency while maintaining a balance with power output, ensuring that the overall performance of the QHE remains robust despite the trade-off considerations when compared to standard measurement based QHE as shown in Fig. \ref{fig:4}.

\begin{figure*}
     \includegraphics[width=0.9\linewidth]{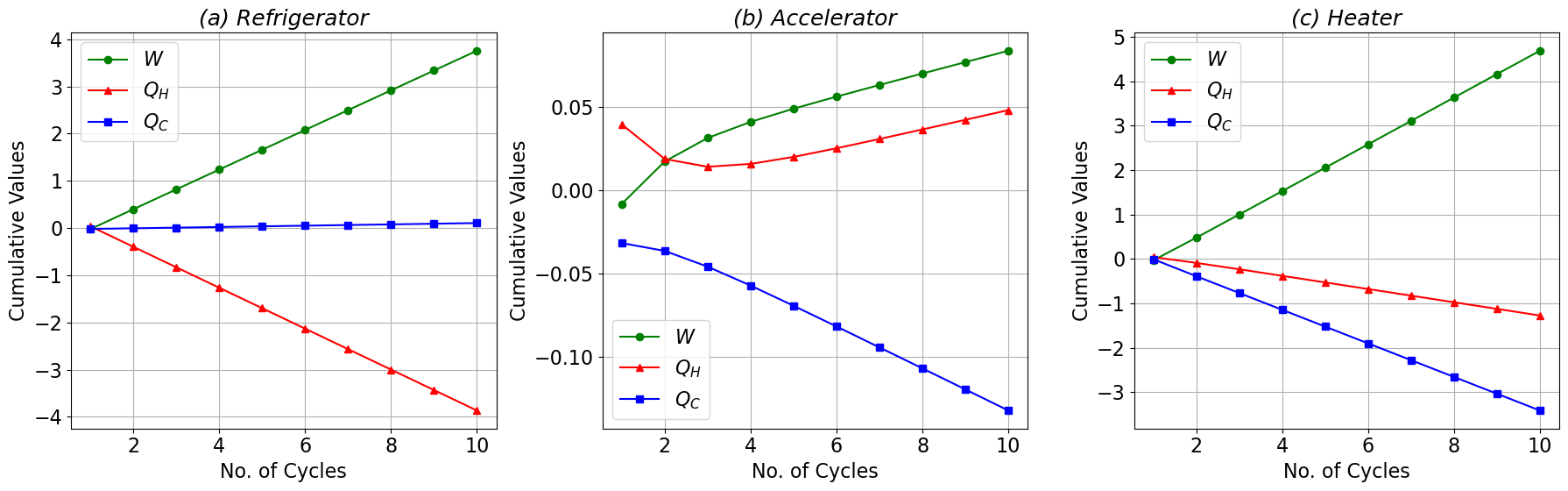}
     \caption{Panels (a), (b), and (c) illustrate the distinct QTMs corresponding to Refrigerator, Accelerator, and Heater operations, as realized in the Quantum Otto Cycle. The parameters used are \(T = 1\) and \(g = 0.75\), with all other parameters and legends identical to those provided in Fig. \ref{fig:8}a.}
     \label{fig:9}
\end{figure*}

\subsubsection{\textbf{Stability Analysis}}

\begin{figure*}[htb]
    \centering
    \begin{subfigure}[b]{0.48\textwidth}
        \includegraphics[width=0.9\textwidth]{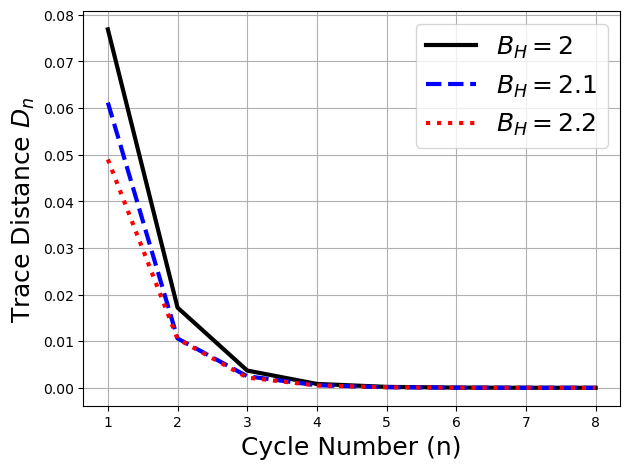}
        \caption{}
        \label{fig:td-b}
    \end{subfigure}
    \hfill
    \begin{subfigure}[b]{0.48\textwidth}
        \includegraphics[width=0.9\textwidth]{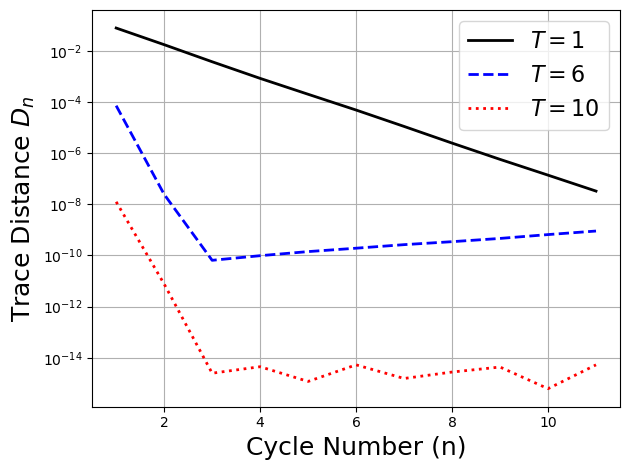}
        \caption{}
        \label{fig:td-t}
    \end{subfigure}
    \caption{Comparison of trace distance evolution with varying parameters B and T. The figure displays the trace distance $D_n = \frac{1}{2} \text{Tr} |\hat\rho_n - \hat\rho_{n-1}|$ as a function of cycle number $n$, with the trace distance plotted for a range of $B_H$ and temperatures $T$. The parameters used here are identical to Fig. \ref{fig:8}. }
    \label{Tracedist}
\end{figure*}

In quantum thermodynamics, a limit cycle is a periodic steady state where the density matrix at the end of each cycle matches the initial state of the next cycle (up to a small numerical error). The trace distance $D_n \rightarrow  0$ (This convergence is quantified by the trace distance, defined as $D_n = \frac{1}{2} \text{Tr} |\hat{\rho}_n - \hat{\rho}_{n-1}|$, which measures the Hilbert-Schmidt distance between consecutive density matrices; here $\hat\rho_n$ is the density matrix after the measurement at the n-th cycle) confirms this. When $D_n \approx 0$, the cycle is effectively closed, and the engine operates reproducibly.  The convergence rate depends on the bath temperature ($T$) and the applied magnetic field ($B_H$), with lower $B_H$ and lower $T$ slowing the process , but all cases eventually reach a near-zero trace distance, indicating robust closure. The measurement on the ancilla at the end of each cycle $(D \rightarrow A)$, followed by its reset to a fixed state (e.g.,$ |00\rangle_{\rm anc}\langle 00|)$, acts as a feedback mechanism. This process projects the system into a state that aligns with the cycle’s dynamics, effectively closing the loop by enforcing consistency between consecutive cycles.

The trace distance tending to zero (as shown in Fig. \ref{Tracedist}) indicates that the measurement and reset process stabilizes the system’s state, as the ancilla’s projection constrains the system’s evolution to a limit cycle. This feedback mimics a cold bath interaction, allowing the system to dissipate entropy and converge.

\subsubsection{\textbf{Entropy Production}}

The entropy production \cite{PhysRevA.101.042106, RevModPhys.93.035008} of a quantum system undergoing a cyclic process, such as the quantum Otto cycle, is a critical measure of irreversibility and is directly tied to the second law of thermodynamics. For the system under consideration, the total entropy production per cycle, denoted $\Sigma$, is given by:
\begin{equation}
\Sigma = \Delta S_{\text{sys}} + \Delta S_{\text{env}} - I,
\end{equation}
where $\Delta S_{\text{sys}}$ is the entropy change of the system, $\Delta S_{\text{env}}$
 is the entropy change of the environment, and (I) represents the mutual information accounting for feedback processes.
The system's entropy change, $\Delta S_{\text{sys}}$ is expressed as the sum of entropy changes across the stages of the cycle:
\begin{multline}
\Delta S_{\text{sys}} = S(\operatorname{Tr}_{\text{anc}}[\hat{\rho}_{B}]) - S(\operatorname{Tr}_{\text{anc}}[\hat{\rho}_{A}]) \\
+ S(\operatorname{Tr}_{\text{anc}}[\hat{\rho}_{D}]) - S(\operatorname{Tr}_{\text{anc}}[\hat{\rho}_{C}]) + \Delta S_{\text{cyc}},
\end{multline}
where $\hat{\rho}_{A}$, $\hat{\rho}_{B}$, $\hat{\rho}_{C}$, and $\hat{\rho}_{D}$ are the density matrices of the system and ancilla at the respective vertices of the quantum Otto cycle. The term $ \Delta S_{\text{cyc}}$ accounts for the entropy change during the reset of the system's state from the post-measurement state $ \hat{\rho}_{A'} := \hat{\rho}_{\text{PM}}$ to the initial state $\hat{\rho}_{A}$
:
\begin{equation*}
\Delta S_{\text{cyc}} = S(\operatorname{Tr}_{\text{anc}}[\hat{\rho}{A'}]) - S(\operatorname{Tr}_{\text{anc}}[\hat{\rho}{A}]).
\end{equation*}
The environmental entropy change, $\Delta S_{\text{env}}$, arises from the interaction with a hot reservoir at temperature (T) and is given by:
\begin{equation*}
\Delta S_{\text{env}} = -\frac{Q_H}{T} = -\frac{\operatorname{Tr}[\hat{H}_B^{\text{sys}} (\hat{\rho}_{C}^{\text{sys}} - \hat{\rho}_{B}^{\text{sys}})]}{T},
\end{equation*}
where $Q_H$ is the heat absorbed from the reservoir, and $\hat{H}_B^{\text{sys}}$
 is the system Hamiltonian at the corresponding stage.
The mutual information term, ($I$), quantifies the information gained through measurement and feedback, defined as:
\begin{equation}
I := I(\text{sys}:\text{anc}) = S(\operatorname{Tr}_{\text{anc}}[\hat{\rho}_{D}]) + S(\operatorname{Tr}_{\text{sys}}[\hat{\rho}_{D}]) - S(\hat{\rho}_{D}).
\end{equation}
This term reflects the correlation between the system and the ancilla, reducing the entropy cost of the system's evolution and enabling higher efficiency without violating the second law.

\begin{figure}[h]
\centering
\includegraphics[width=0.9\linewidth]{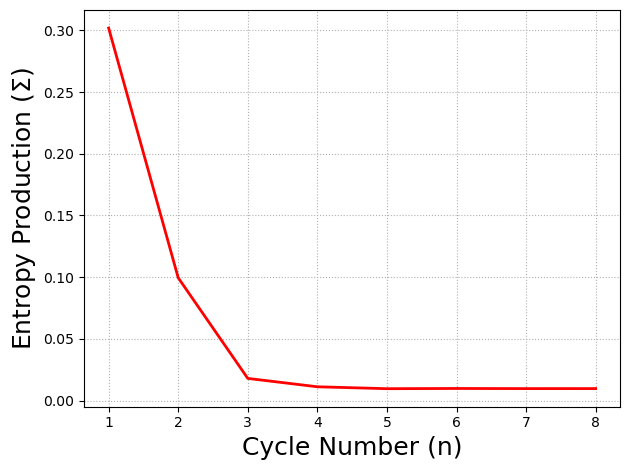}
\caption{Entropy production (\textbf{$\Sigma$}) as a function of the number of cycles. The parameters used here are identical to Fig. \ref{fig:8}.}
\label{fig:Ent Pro}
\end{figure}

The behavior of entropy production over multiple cycles is illustrated in Figure \ref{fig:Ent Pro}, which plots $\Sigma$
 as a function of the number of cycles. This figure highlights the cumulative irreversibility of the process and confirms that the entropy production remains non-negative, addressing concerns about potential violations of the second law.
 
\subsubsection{\textbf{Realisation of different Quantum Thermal Machines}}

Considering the signs of the three distinct energy exchanges during the Quantum Otto Cycle, only four out of the eight possible combinations are permitted by the second law of thermodynamics. This law is expressed as

\begin{equation}
    \beta_1 \langle\Delta E_1\rangle + \beta_2 \langle\Delta E_2\rangle \geq 0,
\end{equation}
where the inverse temperatures \(\beta_1\) and \(\beta_2\) of the interacting thermal baths satisfy the condition \(0 < \beta_1 < \beta_2\). This condition ensures that the hotter bath (with \(\beta_1\)) and the colder bath (with \(\beta_2\)) interact with the working system in a manner consistent with thermodynamic principles, allowing only physically valid energy exchanges. \\

The various modes of operations are given below.\\
\begin{enumerate}
    \item Heat Engine:      $Q_{H} \geq 0, Q_{C} \leq 0, W_{tot} \leq 0$. 
    \item Accelerator:       $Q_{H}\geq 0, Q_{C} \leq 0, W_{tot} \geq 0$. 
    \item Heater:               $Q_{H} \leq  0, Q_{C} \leq  0, W_{tot} \geq 0 $. 
    \item Refrigerator:     $Q_{H} \leq 0, Q_{C} \geq  0, W_{tot} \geq 0 $. 

\end{enumerate}

In this setup, the cold bath is effectively eliminated by selecting an appropriate measurement basis, which is projected onto the ancillary qubits. Different measurement bases are employed to direct the energy flows into and out of the system. We illustrate various operational modes achievable within specific parameter regimes. Operating this QOE with a single thermal bath challenges the validity of the Carnot efficiency limit. By tuning the temperature of the thermal bath, these alternative operational modes become accessible. The refrigeration effect is achieved by projecting the ancilla onto the state \( I_{sys} \otimes|11\rangle_{\text{anc}}\langle11| \), which in turn heats the system. This projection effectively replaces the role of a cold bath in the refrigeration cycle. The other two thermal machines are achieved by measuring the ancilla in the Bell basis $I_{sys} \otimes|\Psi_{anc}^{\pm}\rangle\langle\Psi_{anc}^{\pm}|$ corresponding to the heater and accelerator action as indicated in Fig. \ref{fig:9}.

\section{Discussion}

\subsection{\textbf{Comparison between the Models}}

\begin{figure*}[htb]
    \centering
    % First subfigure
    \begin{subfigure}[b]{0.45\textwidth}
        
        \includegraphics[width=\linewidth]{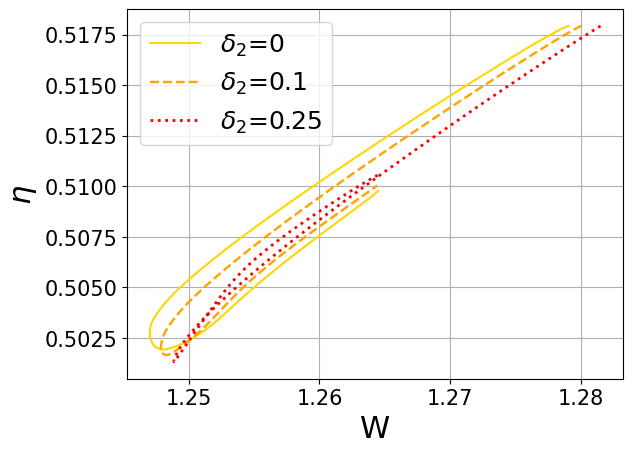}
        \caption{}
        \label{fig:10subfig1}
    \end{subfigure}
    \hfill
    % Second subfigure
    \begin{subfigure}[b]{0.45\textwidth}
        
        \includegraphics[width=\linewidth]{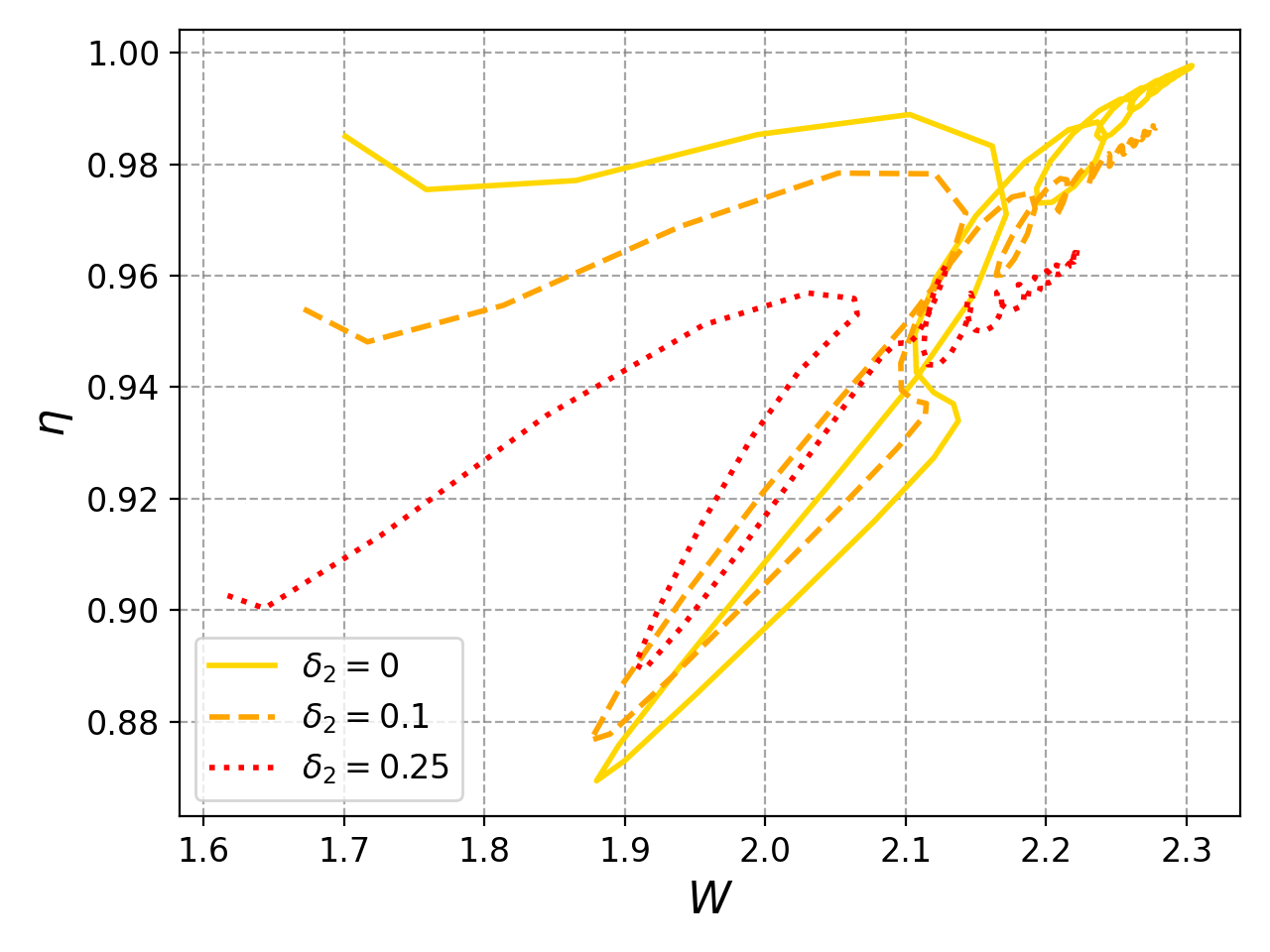}
        \caption{}
        \label{fig:10subfig2}
    \end{subfigure}
    
        \caption{Comparison between models A and B.               (a) Parametric plot between Efficiency ($\eta$) vs $W$ when the measurement is done on the system. The coupling strength between the system qubits $(J_{1})$ is used as a parameter, varied from -2 to 2, which is the peak-to-peak variation as shown in Fig.\ref{fig:3}. (b) Parametric plot between Efficiency ($\eta$) vs $W$ when the measurement is done on the ancilla. $J_{1}$ is varied from 0 to 20 as indicated in Fig.\ref{fig:7a}.}
    \label{fig:main_figure}
\end{figure*}

The Fig. \ref{fig:10subfig1} and \ref{fig:10subfig2} illustrate the parametric variation of efficiency versus work done in the system for measurements performed on the system qubits and ancilla qubits, corresponding to models A and B, respectively. In the case of system-based measurements (model A), the engine's performance slightly exceeds the conventional Otto limit, attributed to the presence of a quantum critical point (QCP). This performance can be further enhanced by tuning the parameter $\delta_{2}$, as shown in Fig. \ref{fig:10subfig1}, where increasing \(\delta_2\) results in a noticeable improvement in efficiency. In practical QHE implementations, decoupling the quantum system from its bath acting as a heat reservoir may not always be feasible, depending on the system-bath architecture. In addition, coupling and decoupling the working system from the bath incurs associated costs. In our model, we initially assume the working system to be completely isolated from its bath during the work-producing stages (AB and CD), preserving their unitary nature, however, we also consider a realistic architecture where the working system remains coupled to its bath throughout the QHE cycle, including stages AB and CD, necessitating the inclusion of energy dissipation to the bath during these stages.

Whereas in the case of ancilla-based measurements, as shown in the efficiency-work plot (Fig. \ref{fig:10subfig2}), the engine outperforms model A, with the efficiency approaching unity for large values of the coupling parameter \(J_1\). Notably, however, an increase in the parametric value of \(\delta_2\) leads to a degradation in the engine's performance. This suggests that while stronger system coupling enhances efficiency, the parameter \(\delta_2\) introduces a competing effect that negatively impacts its overall performance. 

Further demonstrations were carried out for three different
modes of operation of the machine: heat engine, refrigerator,
and heat accelerator. These modes can be observed when
the machine reaches a steady performance limit \cite{PhysRevA.106.032410,PhysRevLett.123.180602}. In the regime $\frac{B_{H}}{B_{L}} <  T_{H} $, it operates as a heat engine, withdrawing heat from the hot bath  ($Q_{H} > 0$), expelling some of it in the cold bath ($Q_{C} < 0$) and producing some useful work ($W < 0$), whereas in the regime $\frac{B_{H}}{B_{L}} > T_{H} $  it works as a refrigerator, consuming work ($W > 0$) in order to extract heat from the cold bath and transfer it to the hot bath ($Q_{C}>0$ and $Q_{H}<0$) and when the projective measurement is done in the ancilla part with $|11\rangle _{anc}$. The measurement on the ancilla using Bell basis $|\Psi_{\pm}\rangle_{anc} = \frac{1}{\sqrt{2}} (|01\rangle \pm |10\rangle) $. The power of engine is not compromised for the QOE in the ancilla assisted measurement as  the power performance of the engine clearly outperforms the standard measurement based engine.

\subsection{\textbf{Experimental Realization}}
Spin chains with Heisenberg type interactions are realizable with advanced experimental platforms, including NMR systems, trapped ion systems, and superconducting qubits. In typical trapped ion setups, the coupling constant can range from several hundred Hz to around 1 kHz, while external magnetic fields can be tuned to a few kHz. For NMR implementations, one example involves the $^{13}\text{C}$ and $^{1}\text{H}$ and nuclear spins in $^{13}\text{C}_{labeled}$, both with spin-1/2. This system offers a relatively high spin-spin coupling constant of $J /2\pi$ = 214.94 Hz, facilitating precise control and measurement of spin interactions. Superconducting circuits \cite{PhysRevA.94.032338,PhysRevB.93.041418} with tunable transmon qubits are widely used by experimentalists as a promising platform for simulating spin chain systems. Each transmon qubit’s frequency is adjustable through applied magnetic flux, allowing precise control of individual qubits. Introducing capacitances, between qubits facilitates an XX-type coupling, while coupling qubit inductances via mutual inductance generates a ZZ interaction. This architecture provides versatile control over qubit interactions, enabling the exploration of quantum spin dynamics and interactions essential for spin chain models.

\section{Conclusion}
We have studied the quantum Otto cycle in a dual pair of spin-1/2 particles with Heisenberg XX interactions, incorporating both inter- and intra-chain couplings which interacts with the single hot bath and we have mimicked the cold bath by using projective measurements in the system and ancilla. In this model, the portion driven by a magnetic field constitutes the working system, while the coupled segment serves as the ancilla. Our analysis spans various time regimes, with a focus on finite-time dynamics in both AFM and FM phases. In the system based measurement we identify QCPs at which efficiency is notably enhanced due to second-order QPTs. This efficiency boost becomes even more significant with an increase in the weak, static transverse field applied to the ancillary part. We further examined the scenario in which the hot bath remains coupled throughout the entire process, resulting in an “\textit{always-on interaction}” model. This continuous interaction diminishes the efficiency of the engine, particularly during extended driving periods. Additionally, we analyzed the impact of varying the measurement basis on the ancilla, revealing that distinct choices yield different types of thermal machines, such as refrigerators, heaters, and accelerators over many number of cycles.

\begin{acknowledgments}
S.R.R would like to acknowledge the financial support provided by the Department of Science and Technology - Science and Engineering Research Board (DST-SERB), which has been instrumental in facilitating this work. Furthermore, we express our sincere gratitude to Dr. Shubhrangshu Dasgupta for his valuable insights and fruitful discussions. \highlight{We thank the reviewers for their thoughtful and constructive feedback, which has significantly improved the quality and clarity of this work.}
\end{acknowledgments}

\onecolumngrid

\appendix
\section{\highlight{Analysis of Energy Level Transitions and Frictional Effects}}
\label{app:transitions}

\highlight{To provide further insight into the engine's performance as discussed in the main text (see Figs.~\ref{fig:4} and \ref{fig:6}), this appendix elaborates on the physical mechanisms governing energy level transitions in the quantum system, particularly under the influence of frictional effects at short times. The power output of the engine, as a function of time $\tau$, is significantly affected by dissipative interactions between the quantum system and its environment, such as a thermal bath. These interactions facilitate energy exchange, driving transitions between the energy levels $E_0$, $E_1$, $E_2$, and $E_3$.}

\highlight{At short times ($\tau < 10$), frictional effects are pronounced, as the system operates in a non-equilibrium state. Initially, at $\tau \approx 0$, the transition probabilities are heavily weighted toward the starting energy level (e.g., $E_0 \to E_0$). As $\tau$ increases, rapid changes occur due to external driving or Hamiltonian variations, with friction providing or absorbing the energy required to bridge energy gaps between levels. This results in significant transition rates, particularly for the $E_0 \to E_0$ probability, which peaks above 0.8, reflecting the system's adjustment to these external influences.}

\begin{figure}[h]
    \centering
    \includegraphics[width=0.8\linewidth]{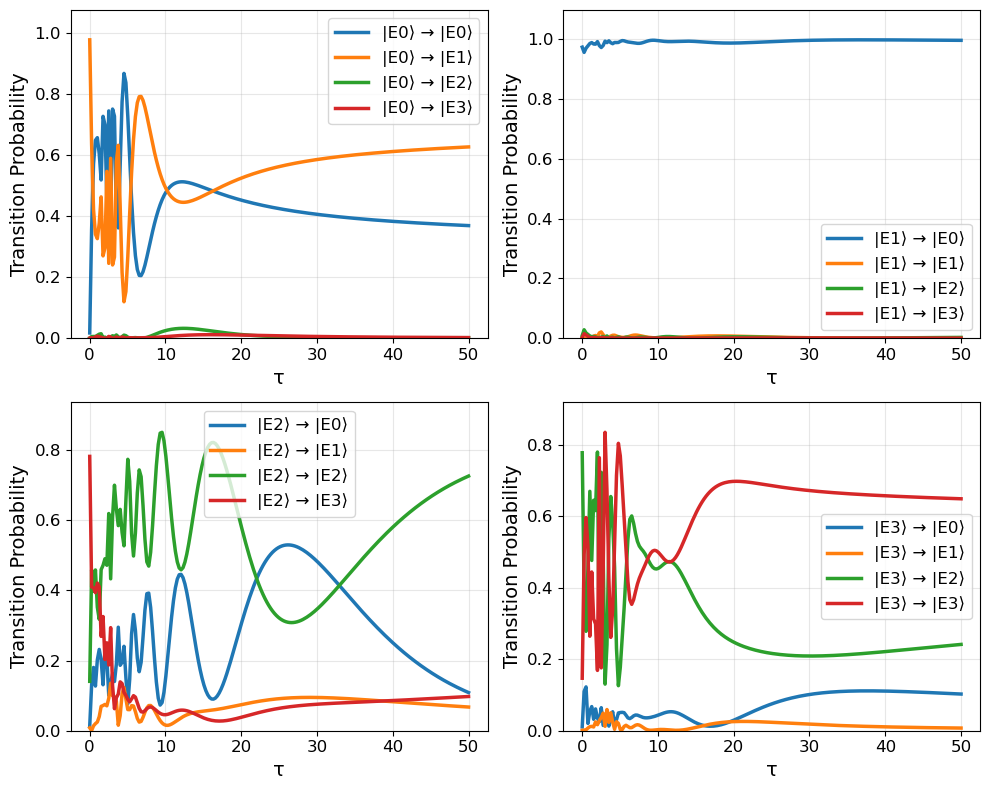}
    \caption{\highlight{Transition probabilities as a function of $\tau$ for different initial states, illustrating the influence of frictional effects on adjacent and non-adjacent energy level transitions. The parameters chosen are $J_1 = 2$, $g = 0.75$, $\delta_{1} = 1$ and all other parameters are same as that of in Fig. \ref{fig:2}}}
    \label{fig:tptau}
\end{figure}

\highlight{The transition probabilities, illustrated in Fig.~\ref{fig:tptau}, reveal that both adjacent and non-adjacent energy level transitions occur, modulated by frictional effects. The following key transitions are observed:}

\begin{itemize}
    \item \textbf{\highlight{Adjacent Transitions:}}
    \begin{itemize}
        \item \highlight{$E_0 \to E_1$: This transition exhibits a strong probability, peaking above 0.8 at $\tau \approx 5$, driven by friction-induced coupling to the next energy level.}
        \item \highlight{$E_1 \to E_0$: A dominant downward transition, stabilizing at probabilities between 0.8 and 0.9.}
        \item \highlight{$E_2 \to E_1$ and $E_2 \to E_3$: These transitions show significant probabilities, with peaks at approximately 0.4 and 0.6, respectively, indicating robust interactions between adjacent levels.}
    \end{itemize}
    
    \item \textbf{\highlight{Non-Adjacent Transitions:}}
    \begin{itemize}
        \item \highlight{$E_0 \to E_2$: This transition, though less probable, reaches 0.1 by $\tau = 50$, indicating a minor but non-negligible pathway across two levels.}
        \item \highlight{$E_2 \to E_0$: A significant non-adjacent downward transition, peaking around 0.6.}
        \item \highlight{$E_3 \to E_1$ and $E_3 \to E_0$: These transitions, spanning two and three levels respectively, peak at approximately 0.6 and stabilize at 0.1, demonstrating that friction enables transitions across larger energy gaps.}
    \end{itemize}
\end{itemize}

\highlight{These observations indicate that adjacent transitions, such as $E_0 \to E_1$, are more probable due to smaller energy gaps and stronger direct coupling. However, non-adjacent transitions, such as $E_2 \to E_0$ and $E_3 \to E_1$, are also significant, particularly at short times when frictional effects dominate. The lower probabilities for transitions like $E_2 \to E_3$ suggest that larger energy gaps reduce the likelihood of such transitions, but the presence of pathways like $E_2 \to E_0$ confirms that friction facilitates a broad range of energy level interactions.}

\twocolumngrid 
\bibliography{main}  % Replace 'references' with your .bib file name
\bibliographystyle{apsrev4-2}  % Standard style for APS journals

\end{document}